\documentclass{elsart5p}
\usepackage{amsmath}
\usepackage{txfonts}
\usepackage[utf8x]{inputenc}
\usepackage{graphicx}
\usepackage[numbers]{natbib}
\def\imagefile#1{{#1.pdf}}
\def\includeplaatje[#1]#2{\includegraphics[#1]{\imagefile{#2}}}%
\def\e#1{\cdot10^{#1}}

\def\lofar{\textsc{lofar}}
\def\lopes{\textsc{lopes}}
\def\kascade{\textsc{kascade}}
\def\coast{\textsc{coast}}
\def\max{\mathrm{max}}

\def\corsika{\textsc{corsika}}
\def\reas{\textsc{reas}}
\def\micro{\ifmmode\mu\else$\mu$\fi}

\def\unitsep{~}
\def\unit#1{\ifmmode\text{\unitsep #1}\else\unitsep #1\fi}
\def\gcm{\ifmmode\textrm{g}/\textrm{cm}^2\else$\textrm{g}/\textrm{cm}^2$\fi}
\let\figurewidth\columnwidth
\def\change#1#2{{#2}}
\begin{document}
\begin{frontmatter}

\title{Prospects for determining air shower characteristics through geosynchrotron emission arrival times}
\author[nijm]{S.~Lafebre\corauthref{cor}},
\corauth[cor]{Corresponding author. Present address: Center for Particle Astrophysics, Pennsylvania State University, University Park, PA 16802, United States. Tel.: +1\,814\,865\,0979; fax: +1\,814\,863\,3297}
\ead{s.lafebre@astro.ru.nl}
\author[nijm,astron]{H.~Falcke},
\author[nijm]{J.~H\"orandel},
\author[fzk]{T.~Huege},
\author[nijm]{J.~Kuijpers}

\address[nijm]
	{Department of Astrophysics, IMAPP, Radboud University, P.O.~Box 9010, 6500GL Nijmegen, The 
Netherlands}
\address[astron]
	{ASTRON, Dwingeloo, P.O.~Box 2, 7990AA Dwingeloo, The Netherlands}
\address[fzk]
	{Institut f\"ur Kernphysik, Karlsruhe Institute of Technology, Campus North, P.O.~box 3640, 76021 Karlsruhe, Germany}

\date{Received $\langle$date$\rangle$ / Accepted $\langle$date$\rangle$}

\begin{abstract}
Using simulations of geosynchrotron radiation from extensive air showers, we present a relation between the shape of the geosynchrotron radiation front and the distance of the observer to the maximum of the air shower. By analyzing the relative arrival times of radio pulses at several radio antennas in an air shower array, this relation may be employed to estimate the depth of maximum of an extensive air shower if its impact position is known, allowing an estimate for the primary particle's species. Vice versa, the relation provides an estimate for the impact position of the shower's core if an external estimate of the depth of maximum is available. In realistic circumstances, the method delivers reconstruction uncertainties down to $30$\unit\gcm\ when the distance to the shower core does not exceed $7$\unit{km}. The method requires that the arrival direction is known with high precision.
\end{abstract}

\end{frontmatter}

%
%
\section{Introduction}
%
%
One of the most important open questions in astroparticle physics is the nature of cosmic-ray particles at the highest energies. At energies exceeding $10^{15}$\unit{eV}, at present, the only practical way to investigate cosmic-ray particles is to register extensive air showers induced by cosmic rays in the atmosphere. In such experiments it is only possible to make statements on the composition of primary cosmic rays based on statistical evaluations. Abundances of primary particle types of an ensemble of air showers are frequently derived by looking at the depth of the shower maximum, i.e.\ the depth at which the number of particles in a shower reaches its maximum.

In recent years, there has been a surge of interest in the detection of extensive air showers by means of the radio emission produced by the shower particles~\citep{2005:Falcke,2005:Ardouin}. This observational technique allows one to look all the way up to the shower maximum, and it has the advantage over detecting the particles themselves at ground level that there is no \change{attenuation}{absorption} of the signal. Several theories explaining the emission mechanism have been proposed \citep{2003:Falcke,2008:Scholten,2008:MeyerVernet}. The former of these explains the observed radio emission from the principle of geosynchrotron radiation, and using a sophisticated model of geosynchrotron emission it was shown that the position of the maximum of inclined showers can be derived from the lateral slope of the electric field strength at ground level~\citep{2008:Huege}.

In this work, we use simulations of air showers and their geosynchrotron radiation to estimate the value of the depth of maximum and the impact position of the shower core. The method developed exploits delays in the arrival time of the signal at different positions on the ground.

%
%
\section{Method}\label{sec:method}
%
%
Detailed distributions of electrons and positrons at different atmospheric depths were obtained from an air shower library~\citep{2007:Lafebre} produced with \corsika\ simulations~\citep{corsika:Heck} and the \coast\ library~\citep{2007:Ulrich}. The library contains air showers initiated by photons, protons, and iron nuclei of energies in the range $10^{16}$ to $10^{20.5}$\unit{eV}, incident from zenith angles up to~$60$º.

\begin{figure}
    \centerline{\includeplaatje[width=\figurewidth]{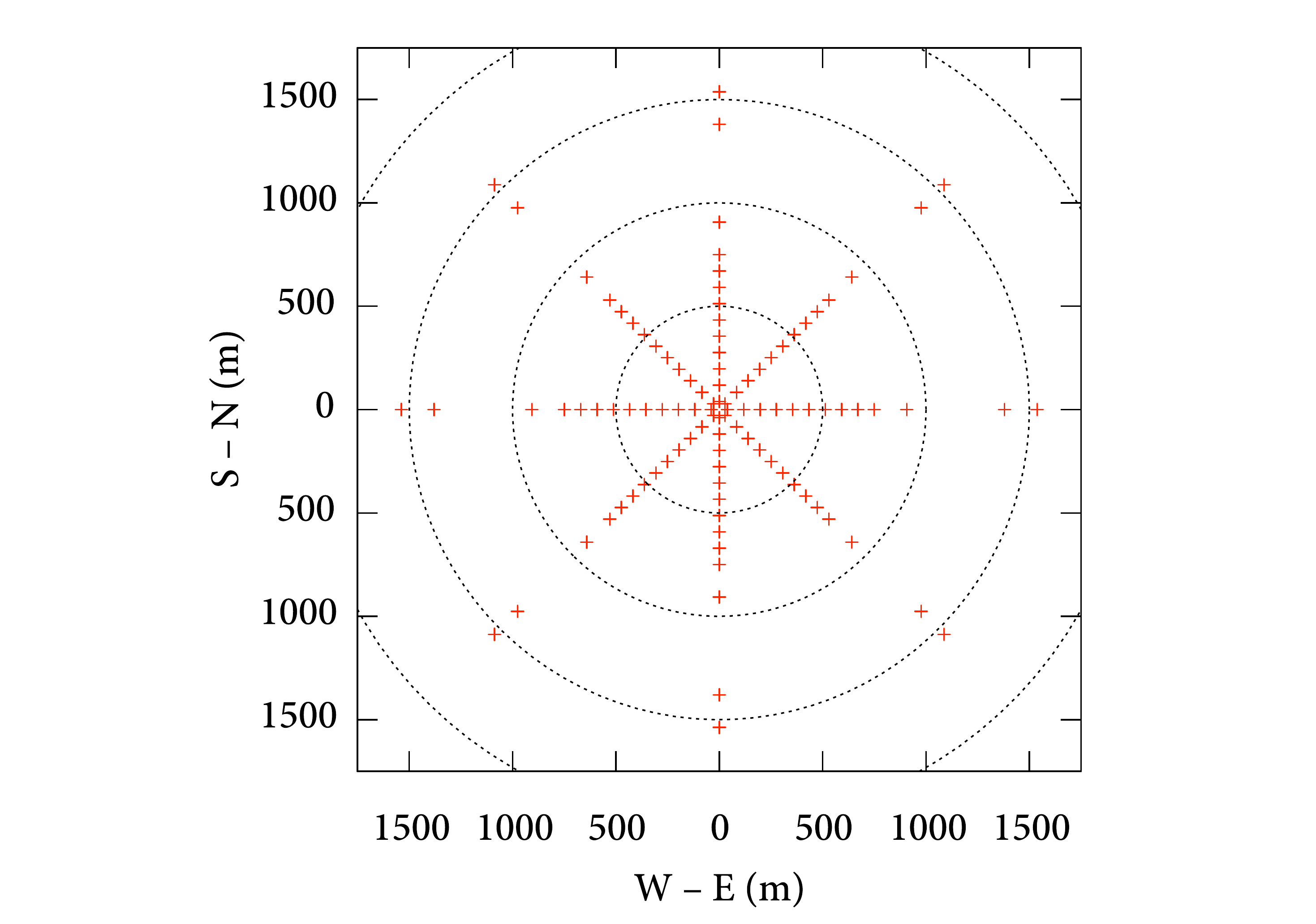}}
    \caption{\change{}{Layout of the virtual array of radio antennas used in the simulations presented in this work. Each marker represents an antenna position.}}
    \label{fig:antennalayout}
\end{figure}
A subset of~$\sim700$ simulations from this library, chosen at random, was used to calculate the radio signal emitted by these airs showers. The \textsc{reas} code version~2.58~\citep{2005:Huege:A&A,2007:Huege} was used to obtain the radio pulses associated with each air shower simulation at an altitude of $100$\unit{m} above sea level. \change{Antennas were placed on a radial grid at distances of $35$\unit{m} to $1500$\unit{m} with intervals of $80$–$300$\unit{m}, with one antenna every $45$º. }{The expected radio signal was calculated for an array of antennas as shown in Fig.~\ref{fig:antennalayout}.}

The magnetic field in all simulations, both \corsika\ and \reas, was taken to match values in northwestern Europe at a field strength of $49$\unit{\micro T} and a declination of $68$º. The height of the detector array was fixed at~$100$\unit{m} above sea level, corresponding to an atmospheric depth of $X\simeq1024$\unit\gcm.

%
%
\section{Parameterization}
%
%
For showers hitting the detector at an angle, one has to compensate for projection effects. Let~$\theta_0$ and~$\phi_0$ be the zenith and azimuth angle at which the primary enters the atmosphere. For a radio antenna a distance~$d$ on the ground away from the shower core in the direction~$\delta$ with respect to the incidence angle~$\phi_0$, the \change{perpendicular distance}{impact parameter}~$r$ is
\begin{equation}\label{eq:showerplane}
    r=d\sqrt{1-\cos^2\delta\sin^2\theta_0}.
\end{equation}
The delay~$\tau$, converted to length units by multiplying with the speed of light in vacuum, is defined as the lag of the peak strength of the radio signal with respect to the arrival time at the shower impact location. It can be written as
\begin{equation}\label{eq:tau}
    \tau = t
          + d\cos\delta\sin\theta_0,
\end{equation}
where~$t(r,\delta)$ is the delay caused by the non-planar shape of the shower front expressed in length units. In the analysis in the remainder of this work, these geometrical compensations have been included.

In the case of a spherical shower particle front, the expected shape of its emitted radio signal is a spherical wavefront as well. The delay~$t$ can then be written in terms of the distance to the center of the sphere~$R$ and the distance from the shower core~$r$ as
\begin{equation}
    t =        \sqrt{R^2+r^2}-R
      \approx  \frac{r^2}{2R},
\end{equation}
where the approximation holds for~$r\ll R$. It was shown previously, however, that the assumption of a spherical shower particle front is unrealistic for large air showers~\citep{2009:Lafebre:Universality}. Therefore, the shape of~$t$ as a function of~$r$ is expected to be different, too.

\begin{figure}
    \centerline{\includeplaatje[width=\figurewidth]{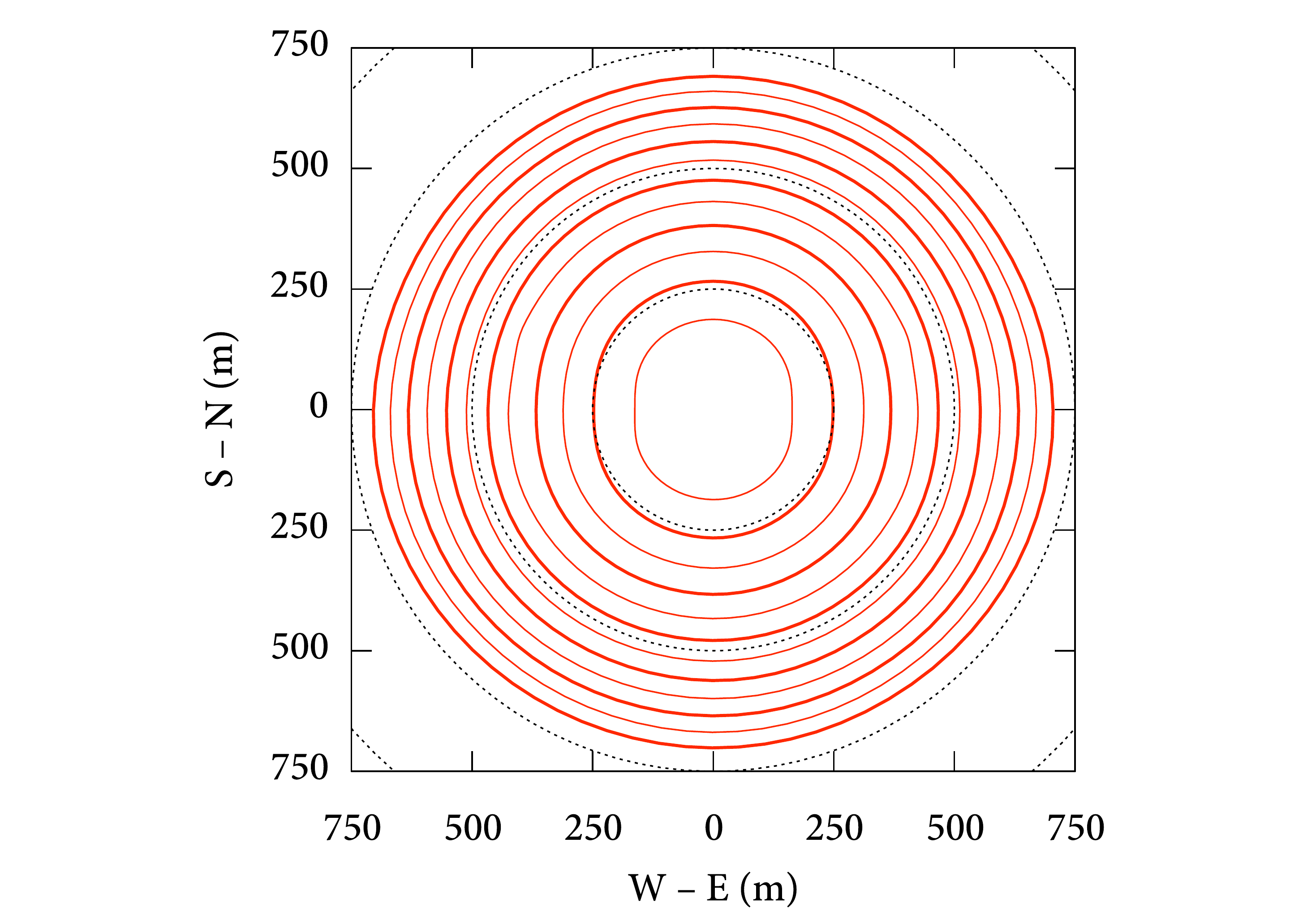}}
    \caption{Radio signal delay for a typical vertical $10^{18.5}$~eV proton shower ($X_\max\simeq780$\unit\gcm). Solid curves represent signal delays (converted to length units)~$\tau$ at intervals of~$5$\unit{m} (thick lines every~$10$\unit{m}). For reference, perfect circles at different distances are also drawn (dotted).}
    \label{fig:frontshape}
\end{figure}
The delay of a radio pulse~$t$ is defined as the lag between a hypothetical plane wave and the actual maximum of the received signal. Fig.~\ref{fig:frontshape} shows a contour plot of the distribution on the ground of this lag for a typical vertical proton shower at $E=10^{18.5}$\unit{eV}, with $X_\max\simeq780$\unit\gcm. The geomagnetic field points north in this figure. Notice the deviation from circularity of the front, which is strongest near the shower core in the east and west directions. This asymmetry results only from radiation processes and is not a consequence of asymmetries in the particle front of the shower, because the distributions used to create the radio shape are cylindrically symmetric by design~\citep{2007:Lafebre}.

Analysis of a set of $\sim700$ showers from photons, protons, and iron nuclei at various energies and incidence angles as described in section~\ref{sec:method} reveals that, to first order approximation, these delays can be described by the parameterization
\begin{equation}\label{eq:parameterization}
    t = R_1^{1-\alpha-1/\beta}r^{\alpha} (R+R_0)^{1/\beta},
\end{equation}
where $R$ represents the distance of the impact location to the shower maximum, which can be translated unambiguously to a value of~$X_\max$. $R_1$ is a scale parameter, the exponent of which was chosen to match the dimension of~$t$ (distance). \change{The distance $R+R_0$ represents the distance from the observer to an imaginary source position from which the air shower originates. This total distance is subdivided into~$R_0$, representing the distance from the point of origin to the shower maximum, and~$R$, which is the distance from the shower maximum to the observer. The value of~$R_0$, which does not greatly influence the overall quality of the parameterization, is fixed at $6$\unit{km}.}{Optimisation of the parameters reveals that the minimum for the $R_0$~parameter is very shallow, and the parameterization can be made to work with a wide range of values for it without appreciable change in quality of the resulting fit. Therefore, $R_0$~was kept at a constant value of~$6$\unit{km} in the final determination of the other parameters to speed up the fit process.}

\begin{figure}
    \includeplaatje[width=\figurewidth]{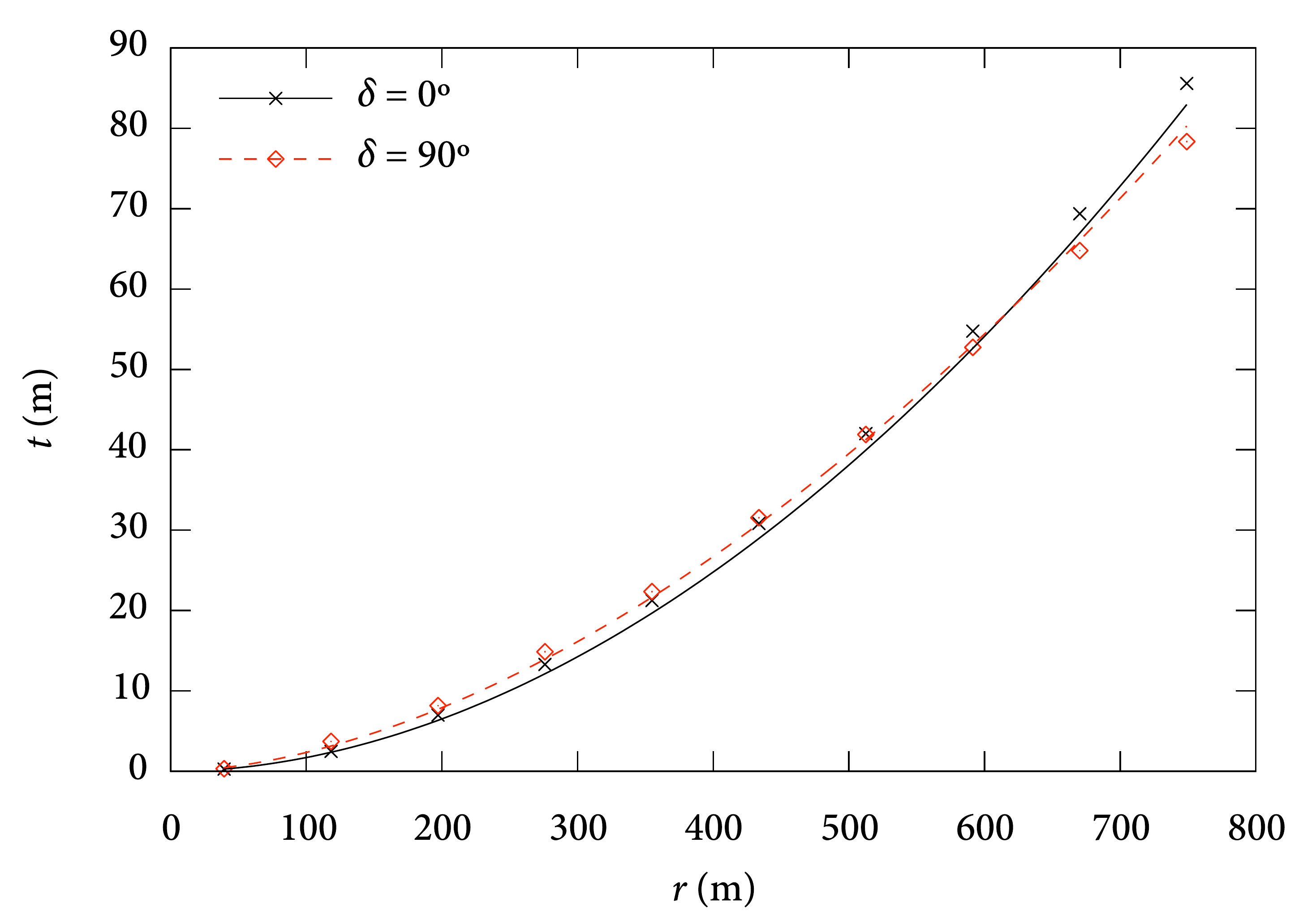}
    \caption{Example of the parameterization presented in \eqref{eq:parameterization} and~\eqref{eq:parameters} for the signal lag for a vertical proton shower at an energy of $10^{20}$\unit{eV} and $X_\max\simeq895$\unit\gcm. The simulated lag at $\delta=0$º and $\delta=90$º is indicated by crosses and diamonds, and their respective corresponding parameterizations are drawn as solid and dashed lines.}
    \label{fig:curvature-example}
\end{figure}
The parameters in the above relation do not depend significantly on either primary energy or zenith angle other than through the respective influences on the depth of the shower maximum. This is not very surprising, because the particle distributions responsible for the radiation do not exhibit any dependence on these parameters either~\citep{2006:Nerling,2009:Lafebre:Universality}. Though the values for~$R_0$, $\alpha$, and~$\beta$ depend on the orientation of the shower with respect to the magnetic field, this dependence is much smaller than the average statistical variation between showers. Therefore, we will restrict the variations in the parameters to a dependence on the angle~$\delta$ only. A fit to the simulated pulse lags in the region $40\unit{m}<d<750\unit{m}$ yields the following overall best-fit parameters:
\begin{align}\label{eq:parameters}
    R_1    &= 3.87  + 1.56 \cos(2\delta) + 0.56 \cos\delta\quad\text{(in km),}\nonumber\\
    \alpha &= 1.83  + 0.077\cos(2\delta) + 0.018\cos\delta,\\
    \beta  &= -0.76 + 0.062\cos(2\delta) + 0.028\cos\delta.\nonumber
\end{align}
The $\cos(2\delta)$ terms in these equations reflect the asymmetries in the east-west versus north-south direction. Note that $\alpha<2$ for all~$\delta$, confirming the non-spherical shape of the wave front. An example of the parameterization is shown in Fig.~\ref{fig:curvature-example}, in which the simulated lags and their corresponding parameterizations are drawn for a vertical proton shower at $10^{20}$\unit{eV} and $X_\max\simeq895$\unit\gcm\ as a function of distance from the shower impact location. Two sets are shown, for $\delta=0$º and $\delta=90$º, respectively.

\begin{figure}
    \centerline{\includeplaatje[width=\figurewidth]{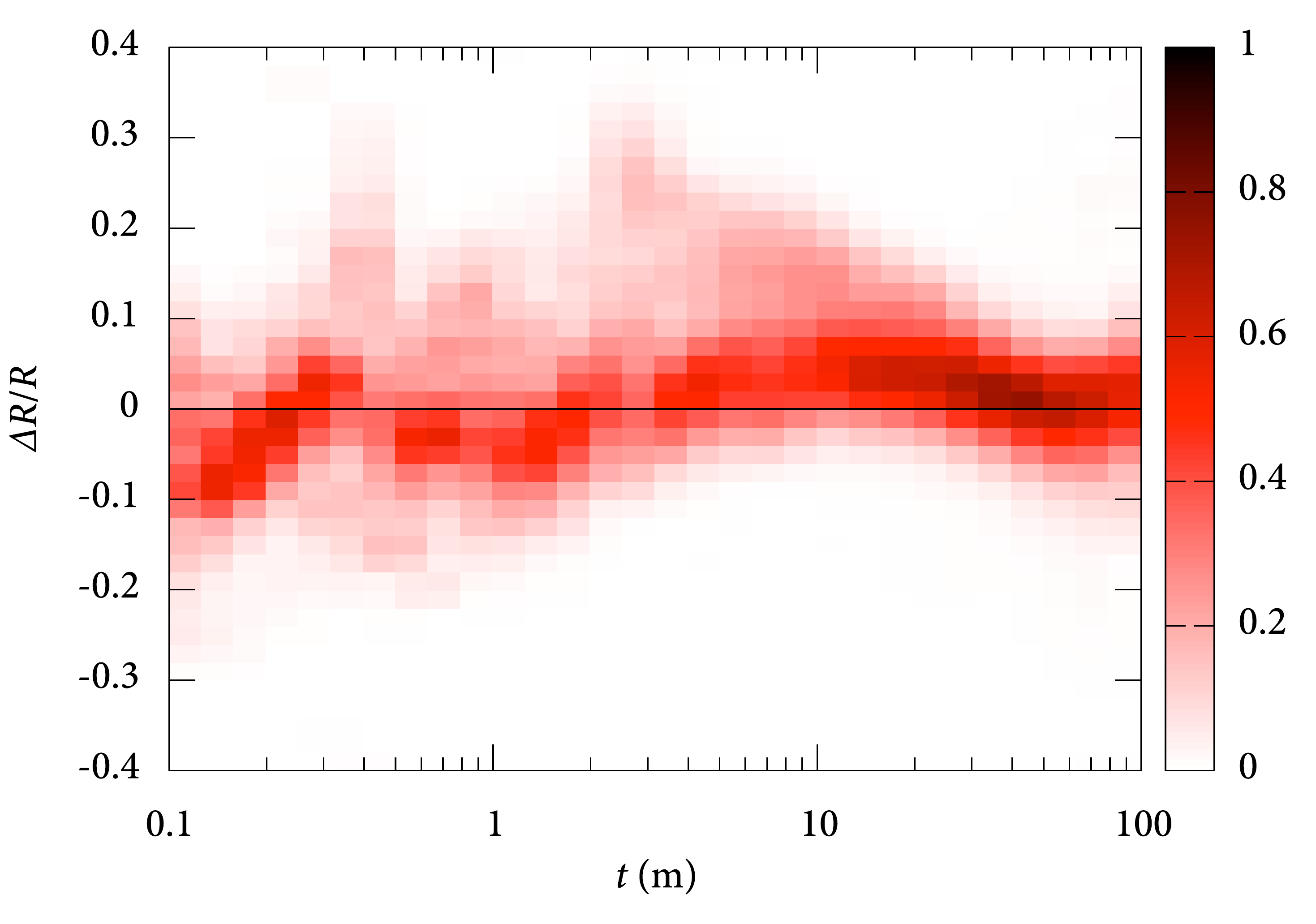}}
    \caption{Relative \change{}{intrinsic} error in the reconstruction of~$R$ as a function of the delay~$t$. Darker areas mark higher numbers of reconstructions. The total amount of colouring is constant for every slice in~$t$; the intensity is in arbitrary units.}
    \label{fig:recerror}
\end{figure}

%
%
%
%
The \change{}{intrinsic} accuracy \change{}{without external error sources} of our parameterization may be assessed from Fig.~\ref{fig:recerror}. This plot shows how the distance to the shower maximum~$R$ as reconstructed from the parameterization in \eqref{eq:parameterization} and~\eqref{eq:parameters} compares to the actual distance as a function of the delay. Note that the figure shows reconstructions of single antennas rather than complete showers: this means that the histogram in this figure is composed of $80\unit{antennas}\times  700\unit{showers}=5.6\e4$ individual reconstructions. It is no surprise that antennas with longer delays of $t>10$\unit{m} produce more accurate reconstructions, since the relative error is smaller there. Even at arrival lags of less than~$1$\unit{m}, however, the standard deviation is less than $10$\unit\% of the actual value.

In a typical array of radio antennas, one can determine the delays~$\tau$ very accurately: using modern equipment, resolutions down to a few~ns can be achieved. We can use the delay values to employ the parameterization in \eqref{eq:parameterization} in two ways: if the position of the shower core is known accurately by scintillator measurements, we can use it to estimate the distance to the shower maximum. If, on the other hand, an estimate for the depth of maximum is available, the position of the shower core can be reconstructed. We \change{will discuss}{explore} these \change{approaches}{possibilities} in detail in the following two sections.

%
%
\section{Determining depth of shower maximum}\label{sec:Xmax}
%
%
\begin{figure*}
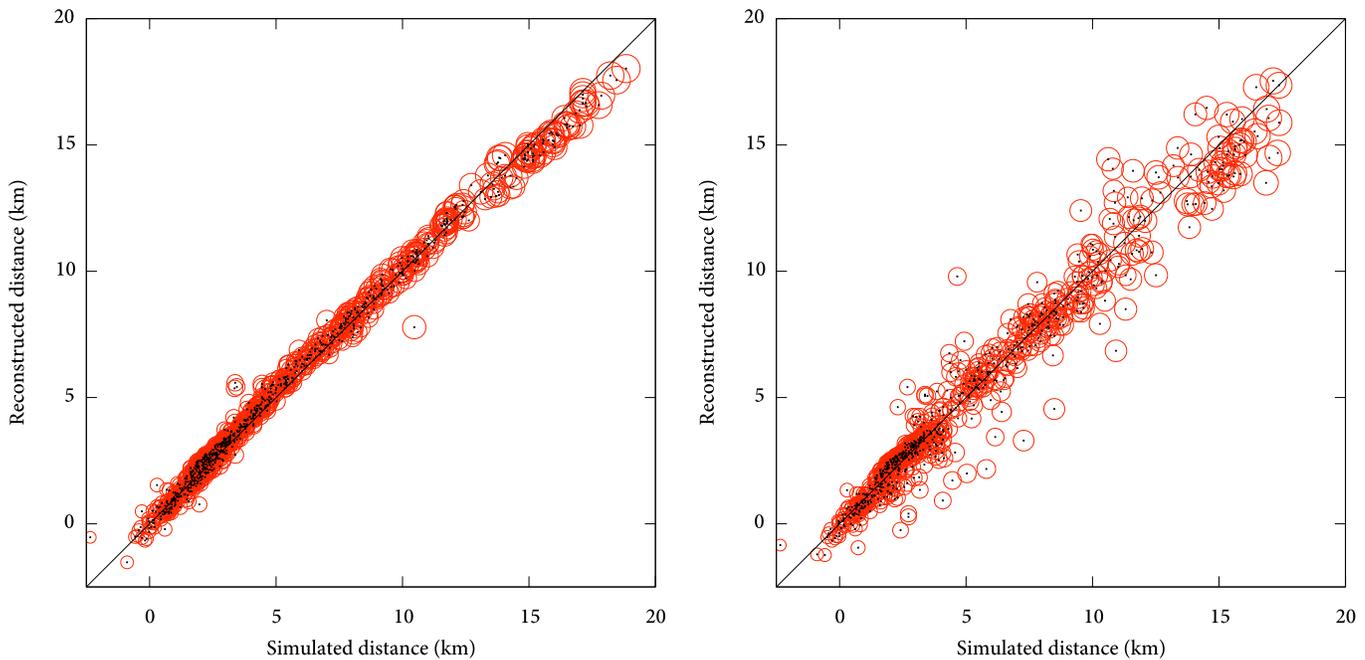

    \includegraphics[width=.5\textwidth]{\imagefile{Reconstructed-distance}}\hfill
    \includegraphics[width=.5\textwidth]{\imagefile{Reconstructed-distance-with-error-estimates}}
    \caption{Scatter plot for~$\sim700$ showers of various species and energies $E>10^{17}$\unit{eV} of simulated values for~$R$ versus the values as reconstructed by the method outlined in the text. Circles around each reconstruction represent error margins of $20$\unit\gcm. The left panel shows the theoretical limit in reconstruction accuracy, while in the right plot realistic Gaussian errors were introduced around the observables in \eqref{eq:reconstruct_R}. \change{}{Note that the distance to the shower maximum extends below zero: these are showers that reach their maximum below the level of the observing radio array.}}
    \label{fig:reconstructed-distance}
\end{figure*}
By rearranging~\eqref{eq:parameterization}, we may write
\begin{equation}\label{eq:reconstruct_R}
    R = R_1^{1-\beta+\alpha\beta} 
        \left( \frac{t}{r^{\alpha}} \right)^{\beta} - R_0
\end{equation}
to reconstruct the distance to the shower maximum. Using this parameterization, the reconstructed distance to the shower maximum is plotted versus the simulated value in the left panel of Fig.~\ref{fig:reconstructed-distance}. Each dot in this plot represents the reconstructed value of~$R$ for one shower event, obtained by taking a weighted average of the reconstructions from the delays in individual antennas. If the antennas are placed on a regular grid, a weight $\propto r^2$ seems justified to match each time delay to its expected relative error, since $\alpha\simeq2$. Our simulated array is denser near the shower core, which was compensated for by multiplying by an extra factor of~$r$, arriving at a total weight for each antenna~$\propto r^3$.

Around each mark in Fig.~\ref{fig:reconstructed-distance} a circle is drawn, the radius of which is the distance corresponding to an atmospheric depth of $20$\unit\gcm\ at the position of the simulated air shower maximum. This value represents the average error for reconstructed $X_\max$~values \change{}{with the Pierre Auger Observatory }using air fluorescence techniques~\citep{2007:Dawson}. The algorithm correctly reconstructs the distance to the shower maximum as simulated, with a standard deviation of $216$\unit{m}. \change{Note that both simulated and reconstructed events extend to negative distances: showers in this region have a maximum that lies below the observation level of the radio antennas.}{Note that for negative distances, the shower maximum lies below the observation level.} By design of he algorithm, correct reconstruction of these \change{events}{negative distances} is possible\change{}{, but} only if the downward distance is smaller than~$R_0$.

\change{So far}{When the uncertainties in Fig.~\ref{fig:reconstructed-distance} are converted to atmospheric depths, we find that the standard deviation of the values for $\Delta X_\max$ is between $15$ and~$20\unit\gcm$ over the full energy range of $10^{16}$–$10^{20}$\unit{eV}. However}, we have \change{}{so far} considered perfect circumstances, assuming exact knowledge of the impact angle and position of the shower axis as well as the delay of the radio pulses. A more realistic picture emerges by introducing some error sources in the reconstruction. For a dense array of radio antennas, such as the \lopes~\citep{2005:Falcke} or \lofar~\citep{2007:Falcke} telescopes, the accuracy in the arrival direction is of the order of~$1.0$º~\citep{2008:Nigl}. A feasible time resolution for determining the maximum pulse height is about~$10$\unit{ns} \change{}{($3$\unit{m}). Because errors in the antenna positions can be reduced to less than $10$\unit{cm} by extended \textsc{gps} measurements, they do not contribute significantly to this uncertainty.} The accuracy in determining the position of the shower core has not been investigated thoroughly yet using radio detection. Therefore, we adopt a typical value from the analysis of the \kascade\ experiment data of~$1$\unit{m}~\citep{2004:Antoni,2005:Glasstetter}. All of the above errors are assumed to follow Gaussian distributions. Additionally, we ensure that the signal is sufficiently strong by demanding a field strength over $180$\unit{\micro V$/$m}, which corresponds to a \change{}{power} signal-to-noise ratio of \change{$10$}{around~$3$} in a rural area~\citep{2008:Huege}.

The right panel of Fig.~\ref{fig:reconstructed-distance} shows the situation when these error estimates are included. The correlation is reduced significantly, which is mainly the result of the uncertainty in the arrival direction of the shower. For very inclined showers in particular this can change the expected delay times dramatically. When the accuracy of the shower impact location is reduced, this mostly affects showers for which the maximum lies at a large distance from the observer. When the error is increased to $5$\unit{m}, for example, hardly any predictions can be made for distances $>10$\unit{km}.

\begin{figure}
    \includeplaatje[width=\figurewidth]{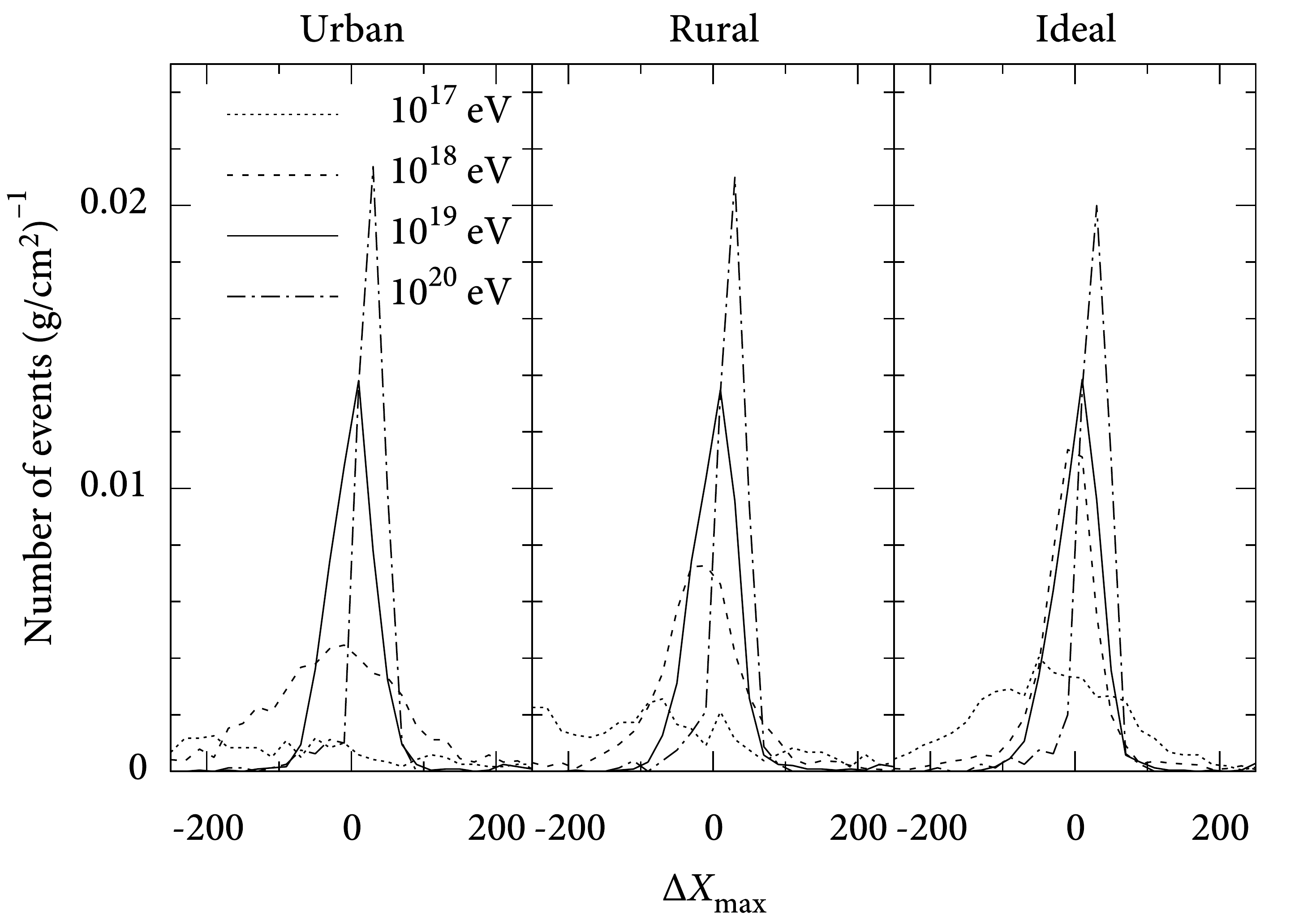}
    \caption{Distribution of residuals for the reconstruction of the depth of maximum for various primary energies. Plots are shown for urban, rural, and ideal noise level scenarios.}
    \label{fig:reconstructed-fraction}
\end{figure}

The distribution of residuals $\Delta X_\max$ (i.e.\ the reconstructed minus the simulated value of the depth of maximum) is shown in Fig.~\ref{fig:reconstructed-fraction} for primary energies between $10^{17}$ and $10^{20}$\unit{eV}. In this plot, a homogeneous detector sensitivity up to zenith angles $\theta<60$º is assumed. Three background noise scenarios are shown: one for an ideal noise level (requiring a field strength $|\vec{E}|>65$\unit{\micro V$/$m} for successful determination of~$t$), one for a rural environment ($|\vec{E}|>180$\unit{\micro V$/$m}), and one corresponding to an urban area ($|\vec{E}|>450$\unit{\micro V$/$m})~\citep{2008:Huege}.

\begin{figure}
    \includeplaatje[width=\figurewidth]{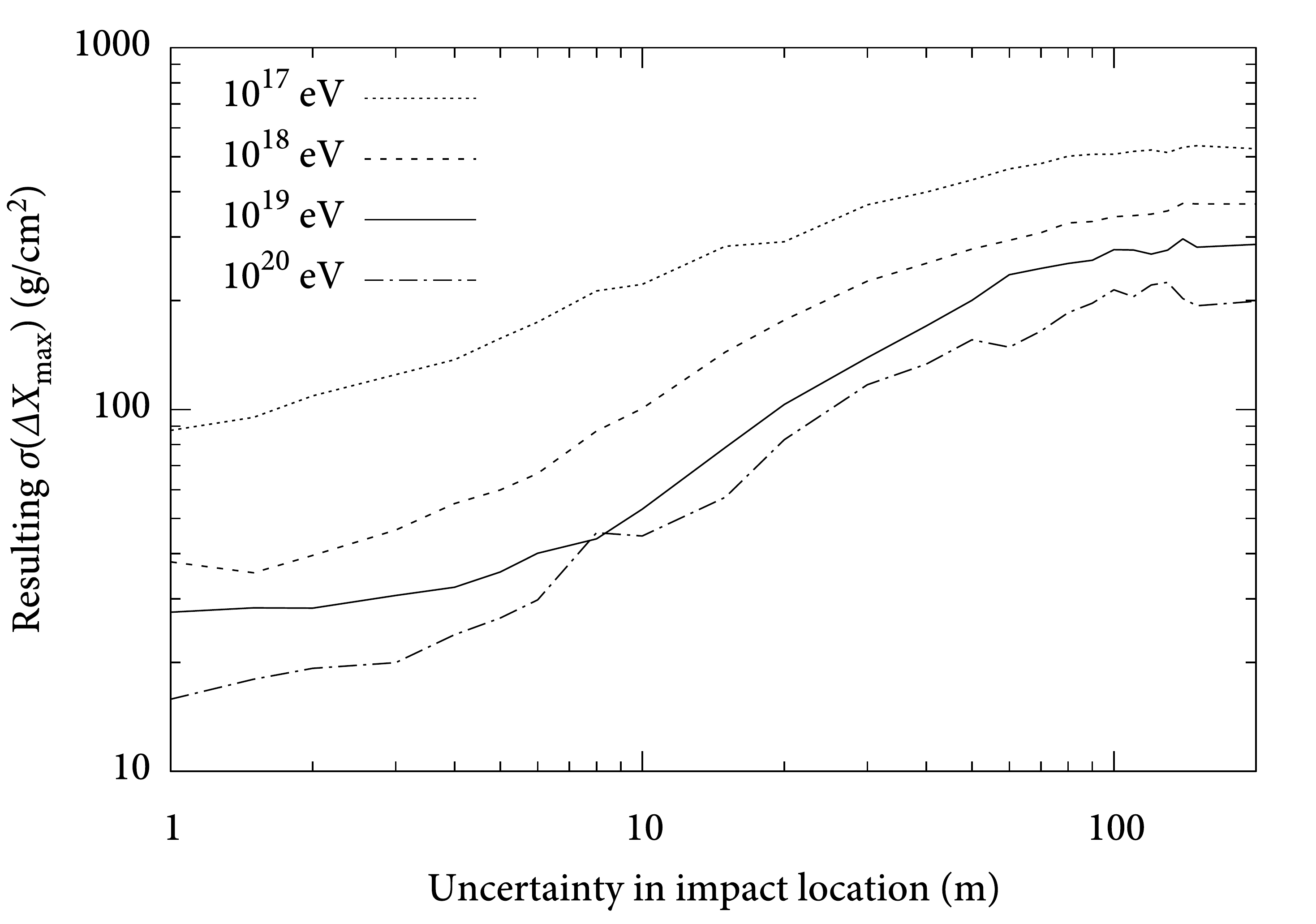}
    \caption{\change{}{Dependence of $\sigma(\Delta X_{\max})$ on the uncertainty introduced in the air shower's impact location for showers of different energies.}}
    \label{xmax-err-vs-pos-err}
\end{figure}

\begin{figure*}
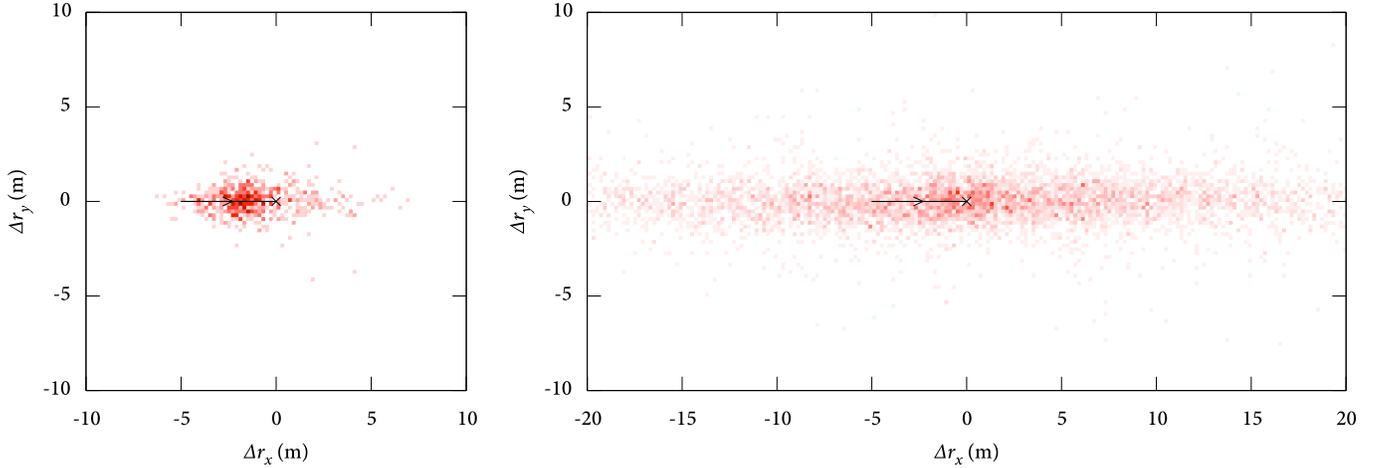

    \includegraphics[width=0.3629\textwidth]{\imagefile{Reconstructed-core-position}}\hfill
    \includegraphics[width=0.6371\textwidth]{\imagefile{Reconstructed-core-position-with-error-estimates}}
    \caption{Density plot for~$\sim700$ showers of various species and energies $E>10^{17}$\unit{eV} of simulated values for the impact location of the shower as reconstructed by the method outlined in the text. The actual position of the core is marked with a cross. Also shown is the arrival direction for slanted air showers (solid line). The left panel shows the theoretical limit in reconstruction accuracy. On the right realistic observational errors were introduced in \eqref{eq:reconstruct_r}. The colour intensity scales linearly with the number of reconstructions at that point.}
    \label{fig:reconstructed-core}
\end{figure*}
From this figure, we observe that the reconstruction accuracy for~$X_\max$ decreases rapidly at low energies. This is because low-energy showers do not occur very deep in the atmosphere on average, raising the distance to the shower maximum, especially in slanted showers. This results in a radiation front with less curvature, necessitating delay measurements further away from the impact location to obtain the same level of reconstruction accuracy. The produced field strength, however, is proportional to the primary energy, decreasing the patch size that is sufficiently illuminated. The combined effect is that it is hard to make correct estimations for the depth of maximum of low energy showers, unless an array at high altitude is employed.

Additionally, the behaviour of the reconstruction accuracy curve at $10^{18}$\unit{eV} in the three scenarios highlights the importance of low background interference levels: the \change{root mean square deviation from the mean}{width of the distribution} decreases \change{from $113$\unit\gcm\ (urban) to~$102$\unit\gcm\ (rural) to~$62$\unit\gcm\ (ideal)}{dramatically} at this energy.
It is also observed that the \change{root mean square value}{distribution width} does not vary much for energies of $10^{19}t$ and $10^{20}$\unit{eV}\change{ at $40$\unit\gcm\ and $30$\unit\gcm, respectively}{}.

\change{}{Fig.~\ref{xmax-err-vs-pos-err} shows the resulting standard deviation in the values for $\Delta X_{\max}$ when the uncertainty in the impact location of the shower is varied. Gaussian error values on other parameters were held constant, and the background noise was $65$\unit{\micro V$/$m}. The values at $1$\unit{m} correspond to the distribution widths in the rightmost panel in Fig.~\ref{fig:reconstructed-fraction}. From Fig.~\ref{xmax-err-vs-pos-err} it is clear that the reconstruction technique employed in this section is better suited for dense arrays, where more accurate impact locations are available.}

If the maximum available distance to the shower core is very small, as would be the case for an array such as \lopes, the fraction of good reconstructions is reduced dramatically. This makes sense, as the shower front shape can no longer be probed accurately. In particular, if the radius of the array decreases to less than~$\sim500$\unit{m}, the amount of useful reconstructions is negligible.

%
%
\section{Determining shower core position}\label{sec:corepos}
%
%
If an estimate for~$X_\max$ (and therefore for~$R$) is available, we can employ~\eqref{eq:parameterization} in an alternative way to estimate values for \change{}{the impact parameter}~$r$, by writing
\begin{equation}\label{eq:reconstruct_r}
    r = R_1^{1+1/\alpha\beta-1/\alpha}\frac{t^{1/\alpha}}{(R + R_0)^{1/\alpha\beta}}.
\end{equation}

In an actual experimental setting, the dependencies of $\alpha$, $\beta$, and~$R_1$ on~$\delta$ need to be taken into account, for example through an iterative fitting procedure for $r$ and~$\delta$. For the sake of simplicity, we will only reconstruct the distance to each antenna here, and we will assume the general direction of the core impact position to be known. This decision is motivated by the fact that the effect on the value of~$r$ caused by variations in~$\delta$ is generally small.

In the theoretical limit, the distribution of reconstructed shower core positions using this method is shown in the left panel of Fig.~\ref{fig:reconstructed-core}. The colouring in this plot shows the amount of reconstructions at a certain position relative to the actual core impact location. The true position is at the origin, indicated by a cross. The arrival direction of inclined showers is always from the left, as indicated by the arrow. Note that the elongated structure of the reconstruction distribution is not a projection effect from inclined showers: we have already compensated for this by the transformation to the shower plane through~\eqref{eq:showerplane}. Instead, the feature is a systematic error intrinsic to the reconstruction algorithm. For a shower incident from the south, for example, the parameterized form is not symmetric in the north-south direction, but it is in the east-west direction. This effect is also responsible for the slight offset of nearly~$-2$\unit{m} in the $\hat x$~direction.

Theoretically, the systematic offset could be reduced and possibly even removed entirely by refining the parameterization in \eqref{eq:parameterization} and~\eqref{eq:parameters}. There is little gain in this exercise, however, when a more realistic reconstruction estimate is made. This is clarified in the right panel of Fig.~\ref{fig:reconstructed-core}, where again some error sources were introduced. The error in the arrival direction is again~$1.0$º, and a Gaussian uncertainty of $20$\unit\gcm\ in the value of the shower maximum is assumed, corresponding to a typical error in~$R$ of $200$–$250$\unit{m}. Clearly, the offset mentioned earlier is entirely swamped by the deviations induced by the uncertainties. The substantial difference in reconstruction accuracy between the $\hat x$~and $\hat y$ direction results directly from the uncertainty imposed on~$\theta_0$: even a small deviation of the zenith angle will make a noticeable difference in the obtained value for~$t$ from~\eqref{eq:tau}.

Similar to the determination of~$X_\max$, the average error increases drastically when the radius of the array is smaller than $500$\unit{m}. The error does not increase significantly, however, when the minimum distance is set to $300$\unit{m}. This is slightly counterintuitive, but it is again related to the accurate probing of the shower front shape. Of course, the requirement remains that the arrival delay at the impact location is known to $10$\unit{ns} or so.

%
%
\section{Discussion}
%
%
\begin{figure}
    \includegraphics[width=\figurewidth]{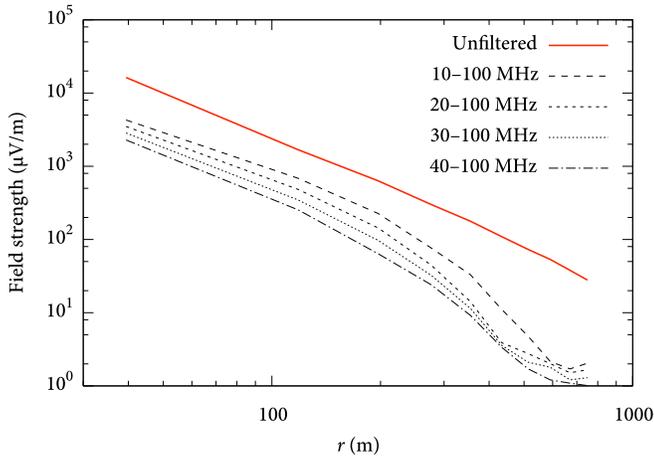}
    \caption{\change{}{Effects on the maximum field strength to the east of a vertical $10^{18}$\unit{eV} shower arising from applying a rectangular filter to the raw radio pulse. From top to bottom, the unfiltered pulse is shown, and the same pulse with filters of $10$, $20$, $30$, and $40$–$100$\unit{MHz} applied.}}
    \label{fig:filtered}
\end{figure}
The analysis in this work on the relative delays of geosynchrotron emission from extensive air showers was performed on the raw, unfiltered pulse shape. In real experiments, however, the antennas used are bandwidth-limited, which will be reflected in \change{}{the measured field strength of the pulse.} \change{The effect is negligible for close antennas ($r<300$\unit{m}), but for remote antennas it will become important, as the pulse is much broader in these regions. In particular, this may affect antennas which clip frequencies below $\sim 40$\unit{MHz}. }{This is illustrated in Fig.~\ref{fig:filtered}, which shows the maximum value of the measured field strength when different rectangular filters are applied to the raw pulses. When frequencies below $40$\unit{MHz} are clipped, the field strength is around 10\% of the unfiltered value over the entire distance range.}

Another effect that has not been investigated is that of the observer's altitude: in our simulations, this height was fixed at $100$\unit{m} above sea level. We do not anticipate a significant change of the parameterization or its parameters, however. This can be inferred from the fact that the description is valid independent of zenith angle. Changing this angle is comparable to varying the observer's altitude.

Though a deviation from a planar wave is indeed observed in \lopes\ measurements~\citep{2005:Falcke}, at only $200$\unit{m} the array is too small to benefit from the theoretical knowledge of the shape of the radio pulse front. There are currently two other experiments under construction, however, that could make use of the technique outlined in this work. One of these is the initiative in which radio antennas inside the Pierre Auger Observatory~\citep{2004:Abraham} will be erected~\citep{2007:Berg}. Such an array could use the method in Sect.~\ref{sec:corepos} to increase the accuracy of the estimated core impact position, since its reconstruction error for the surface detectors is in excess of~$100$\unit{m}. A precise estimate for~$X_\max$ would have to be provided by the fluorescence detectors. The planned spacing of radio antennas is $150$–$375$\unit{m}, which would allow an accuracy in the reconstruction of around~$30$\unit{m} if the core lies within the radio array. \change{}{Using the Auger array for the metod outlined in Sect.~\ref{sec:Xmax} would probably not be possible, as the uncertainty in the reconstructed core position of around $150$\unit{m} would wash out any sensitivity of the algorithm to the shower maximum.}

Another possible experiment is the \lofar\ telescope~\citep{2007:Falcke}, which consists of a dense core of approximately~$2$\unit{km} in diameter, with groups of~$48$ radio antennas every few hundred meters. Its size and spacing make this setup ideally suited to determine~$X_\max$ using the method outlined in Sect.~\ref{sec:Xmax}. \change{At present, no hybrid detection method is available for \lofar, however, so the shower core position has to be determined too by radio methods, making the estimates and their errors dependent on one another.}{The shower core position, which would have to be known to apply the method, could be obtained in several ways. First of all, there is a small scintillator array coincident with the \lofar\ core, allowing an independent measurement of this quantity. Alternatively, pulse shape and lateral slope of the radio signal could be used to get an estimate for the core position~\citep{2006:Ardouin}. It is assumed that reconstruction with a dense radio array such as \lofar, which places antennas at distances of the order of 10\unit{m}, will be on a par with the precision level of scintillator arrays.}

\change{}{Pulse shape and lateral slope also contain additional information about the value of $X_\max$, with precisions of up to $16$\unit\gcm\ \citep{2008:Huege}. Ideally, one would combine the two methods in a single fit to obtain the best possible reconstruction accuracy.}

%
%
\section{Conclusion}
%
%

Through detailed simulations of air showers and their geosynchrotron radio emission, we have derived an empirical relation between the relative delay of the radio pulse emitted by the air shower front and the atmospheric depth of the shower maximum. By analysis of the radio pulse arrival delays in radio antennas in an array of low-frequency radio antennas, this relation can be used to estimate the depth-of-maximum if the impact position is known or vice versa.

We have confirmed that both methods work in principle, with no information other than radio signal delays used in the reconstruction. When the algorithm is tested under realistic conditions, however, the accuracy of the method is reduced. In the case of determining the shower maximum, reconstruction down to a useful confidence level is possible only for shower maxima up to~$\sim7$\unit{km} away, and only if the shower core impact position is known down to a few meters. When the parameterization is used to derive this position, the critical quantity is the accuracy in the zenith angle of the shower, which needs to be significantly less than a degree to reconstruct the shower impact location to an accuracy of $10$\unit{m} at high inclinations up to~$60$º.

%
%
\begin{ack}
This work is part of the research programme of the `Stich\-ting voor Fun\-da\-men\-teel Onder\-zoek der Ma\-terie (\textsc{fom})', which is financially supported by the `Neder\-landse Orga\-ni\-sa\-tie voor Weten\-schap\-pe\-lijk Onder\-zoek (\textsc{nwo})'. T.~Huege was supported by grant number VH-NG-413 of the Helmholtz Association.
\end{ack}
%
%
\bibliographystyle{unsorted}
\begin{small}

\end{small}

\end{document}